# Rare Association Rule Mining for Network Intrusion Detection

[1]Hyeok Kong , [2]Cholyong Jong and [3]Unhyok Ryang

[1,2]Faculty of Mathematics, Kim Il Sung University, D.P.R.K

[3]Information Technology Research Institute, Kim Il Sung University, D.P.R.K

[1]hyeok_kong@yahoo.com ,[2]jcy0820@yahoo.com and [3]liangzhengai@yahoo.com

## Abstract

In this paper, we propose a new practical association rule mining algorithm for anomaly detection in Intrusion Detection System (IDS). First, with a view of anomaly cases being relatively rarely occurred in network packet database, we define a rare association rule among infrequent itemsets rather than the traditional association rule mining method. And then, we discuss an interest measure to catch differences between interesting relations and uninteresting ones, and what interest there is, and develop a hash based rare association rule mining algorithm for finding rare, but useful anomaly patterns to user. Finally, we define a quantitative association rule in relational database, propose a practical algorithm to mine rare association rules from network packet database, and show advantages of it giving a concrete example.

Our algorithm can be applied to fields need to mine hidden patterns which are rare, but valuable, like IDS, and it is based on hashing method among infrequent itemsets, so that it has obvious advantages of speed and memory space limitation problems over the traditional association rule mining algorithms.

## 1. Introduction

Today, with a rapid advance of information technology, a large quantity of data are gushed out with a big spurt every day and every minute, and the vast data cannot be possibly analyzed by human abilities. In an information age, it is a key to use valuable information in making a determination, so that we cannot make advance without data mining technology any more.

Strictly speaking, data mining is a process of discovering valuable information from large amounts of data stored in databases, data warehouses, or other information repositories. This valuable information can be such as patterns, associations, changes, anomalies and significant structures. Specially, the research on association rule mining has widely done in many application domains during the past decade.

A typical example of association rule mining is "market basket analysis", which aims at finding relations among itemsets in transaction database. This finding of relations is based on the "frequent itemsets" in transaction database. As the origin of association rule mining problem is the market basket analysis, the traditional algorithms including Apriori and FP-growth for association rule mining, focused to find efficiently the frequent itemsets. There are quite a number of "infrequent itemsets" than frequent-ones in database for market basket analysis, and these algorithms can be improved, dealing with such infrequent item- sets as little as possible.

A significant drawback of the traditional approach to association rules is the large number of rules that are generated and processed. Even a small database often generates several thousand rules. In addition, potentially interesting infrequent itemsets are discarded a priori by the definition of the association rules.

However, network packet database for IDS, which is another recent application of data mining, has more normal activities than abnormal ones of users, so that there are a number of frequent itemsets than infrequent ones in it, the algorithms for IDS can be improved, finding the associations among infrequent itemsets rather than frequent ones. That is the purpose to discover rare association rules of small frequencies, but strong interest in IDS.

Consequently, we can find the differences in two typical application examples of so called the market basket analysis and the intrusion detection as follows.

(1) While the database for market basket analysis is a transaction database in which each transaction has different length (i.e. the number of data items in a transaction), the database of for intrusion detection is a relational database of which record length is same.

(2) In a realistic case, there can be many hundreds or even many thousands of products (data items) in database for market basket analysis. In contrast to this, network audit databases face tens of attributes.

(3) For market basket analysis, an association rule is the implication $X \rightarrow Y$, where X and Y are itemsets like $\{I1, I2, \ldots, In\}$. But, for intrusion detection, X and Y are itemsets like $\{f_1=q_1, f_2=q_2, \ldots, f_n=q_n\}$, where $f_k (k=1, 2, \ldots, n)$ is item name (field name) and $q_k (k=1, 2, \ldots, n)$ is a value of item.

From above consideration, first, we formulate a rare quantitative association rule in the relational database, through earlier association rule mining in transaction database. And then, we develop a hash based rare association rule mining algorithm for IDS.

## 2. Related works

For more than a decade, researches on association rule mining have attracted a huge interest from the data mining communities. Many advances in association rule mining have been proposed in recent years, including more efficient algorithms to process association rules, new data structures to speed up processing, new compression techniques to overcome the memory limitation problem, and so on.

However, most of the existing research on association rules has been focusing on establishing common patterns and rules; these are patterns and rules based on the majority, some of which may be either obvious or irrelevant. [1,2,3] Unfortunately, not enough attentions have been given to mining rare association rules; these are outlier rules and patterns.

Rare association rules are critically important in many application domains such as intrusion detection and medical diagnosis, they need to discover unusual patterns, which cannot be easily discovered by the traditional association rule mining algorithms.

In [4], were proposed two novel algorithms called MBS and HBS for efficient discovery of association rules among infrequent items. In matrix based scheme, say MBS, the index function, say I(x, y), was employed to identify the index values for any $k$-infrequent itemsets in a given transaction directly.

They also proposed hash based scheme, say HBS, to overcome the drawback of MBS, that is the limitation of memory space, discussed quantitative association rules in transaction database, but in relational ones, and did not closely combine intrusion patterns with rare association rules.

In [5], were also considered anomalous rules to obtain new kinds of association rules that represent deviations from common behaviors and discussed the importance of the rare association rule mining problem in agriculture and medicine. And they discussed a new set of anomalous association rules that can identify several kinds of hidden patterns, including those that may occur infrequently. But, they didn't introduce an interest measure to separate several kinds of hidden patterns, so their method is not practical.

In this paper, we discuss a practical rare association rule mining problem that is critically arisen from application fields such as intrusion detection, needs to discover useful patterns which rarely occur. To do this, we formulate an association rule among frequent itemsets and a rare association rule among infrequent itemsets, to clear difference of them. Especially, we introduce a concept of an interest measure to catch differences between interesting relations and uninteresting ones, and what interest there is, and develop a hash based rare association rule mining algorithm for finding rare and useful anomaly patterns.

And, we define a quantitative association rule in relational database and consider a practical algorithm to mine rare association rules from network packet database using an example.

## 3. Definitions of rare association rules

Most of association rule mining algorithms employ minimum support and confidence thresholds to find interesting rules. Although these two parameters prune many associations discovered, many rules that are not interesting to the user may still be produced.

For example, Table 1 shows the purchase data of Soy and Salt in a supermarket. Here, the support for rule "Soy→Salt" is 20%, which is fairly high. The confidence is the conditional probability that a customer buys Salt, given that he/she buys Soy, i.e. $P(Soy \wedge Salt) / P(Soy) = 20/25 = 0.8$, or 80%, which is also fairly high. At this point, we may conclude that the rule "Soy→Salt" is a valid rule.

However, "Soy→Salt " is misleading since the probability of purchasing Salt is 95%, which is even larger than the confidence 80%. In fact, Soy and Salt are negatively correlated since the purchase of one of these items actually decreases the likelihood of purchasing the other.

Table 1. Purchase of Soy milk and Cow milk in a supermarket

|  | salt | ¬ salt | Σrow |
|---|---|---|---|
| soy | 20 | 5 | 25 |
| ¬ soy | 70 | 5 | 75 |
| Σcol | 90 | 10 | 100 |

The above example indicates the weakness of association rule mining using support and confidence. That is, if the occurrence of antecedent does not imply the occurrence of consequent, rule can be misleading.

Hence, in this paper we propose an alternative method for finding interesting relationships between data itemsets based on correlation.

Given a database, let A be an itemset in it and the support, say supp(A), supp(A) = p(A). Then, a statistical definition of dependence for the itemsets X and Y in the database is as follows.

$$\text{Interest}(X, Y) = supp(X \cup Y)/(supp(X)supp(Y)).$$

If $supp(X \cup Y)/(supp(p(X)p(Y))) = 1$, then $supp(X \cup Y) = supp(X)supp(Y)$, i.e. Y and X are independent.

If $supp(X \cup Y)/(supp(p(X)p(Y))) > 1$, then $supp(X \cup Y) > supp(X)supp(Y)$, i.e. Y is positively dependent on X.

If $supp(X \cup Y)/(supp(p(X)p(Y))) < 1$, then $supp(X \cup Y) < supp(X)supp(Y)$, i.e. Y is negatively dependent on X. (or ¬ Y is positively dependent on X).

That is, if the value of Interest is 1, then Y and X are independent and there is no interest in the rule X→Y. It means if antecedent and consequent are independent, then the rule has no interest. If the value of Interest is not 1, i.e. $|supp(X \cup Y)$ − $supp(X)supp(Y) | \geq$ min_interest, then we can consider the itemset $X \cup Y$ has potentially interest.

**Definition 1**. Let I={$I_1$, $I_2$, …, $I_n$} be a set of all items in a transaction database, X,Y⊆I be itemsets, X∩Y=∅ , supp(X)≠0, and supp(Y)≠0. Also, the thresholds: min_supp, min_conf and min_interest > 0 are given by users or experts. Then, the rule X→Y can be extracted as *a valid rule of interest* or *a association rule* if

(1) supp(X∪Y) ≥ min_supp,

(2) |supp(X∪Y)−supp(X)supp(Y) | ≥ min_interest,

(3) conf(X→Y) = supp ( X∪Y) / supp (X) ≥ min_conf.

Here, condition (1) ensures that X and Y are frequent itemsets, condition (2) ensures that X→Y indicates a positive association rule in the case of Interest >1 and a negative association rule in the case of 1 > Interest > 0, and condition (3) describes the strength of the relation.

**Definition 2**. Let $I=\{I_1, I_2, \ldots, I_n\}$ be a set of all items in a transaction database, $X,Y \subseteq I$ be itemsets, $X \cap Y = \emptyset$ , supp(X)≠0, and supp(Y)≠0. Also, the thresholds: min_supp, min_conf and min_interest > 0 are given by users or experts. Then, the rule X→Y can be extracted as *a valid rare rule of interest* or *a rare association rule* if

(1) supp(X) ≤ min_supp, supp(Y) ≤ min_supp,

(2) |supp(X∪Y)−supp(X)supp(Y)| ≥ min_interest,

(3) Interest(X, Y) > 1,

(4) conf(X→Y) = supp ( X∪Y) / supp (X) ≥ min_conf.

Here, condition (1) ensures that X and Y are infrequent itemsets, condition (2) ensures that X→Y is a rule of interest. From condition (3), the rules are restricted to positive ones, and condition (4) describes the strength of the relation.

**Example:** Let min_supp=0.5, min_conf=0.5, and min_interest=0.05. Also, we suppose that supp(X)=0.4, supp(Y)=0.3, and supp(X∪Y)=0.3. We can take

(1) supp(X)=0.4 < 0.5, supp(Y)=0.3<0.5,

(2) |supp(X∪Y) −supp(X)supp(Y)| =|0.3−0.4*0.3 |= 0.18>0.05,

(3) Interest(X, Y) = supp(X∪Y)/ (supp(X)supp(Y))　=0.3/(0.4*0.3) = 2.5> 1,

(4) conf(X→Y) = supp ( X∪Y) / supp (X) = 0.3/0.4 = 0.75 ≥ 0.5.

Hence, rule X→Y is extracted as a rare associative rule.

Now, let R be a relational database; $f_1$, $f_2$, ..., $f_n$ be field names of R; and $q_{jk}$ ($j=1$, ..., n; $k=1$, ..., $j_m$) be values of the field $f_j$. Each record $R_i$ in R is separated by a identifier ID and it is represented as $R_i=\{f_1=q_{1i},\ f_2=q_{2i},\ ...,\ f_n=q_{ni}\}$. And every itemset in R is represented as $I=\{f_j = q_{jk}/\ j=1,\ ...,\ n;\ k=1,\ ...,\ j_m\}$.

$X = \{f_1 = p_1,\ ...,\ f_k = p_k\} \subseteq I$ is called *a quantitative itemset*. Here, $f_1$, ..., $f_k$ are field names and $p_1$, ..., $p_k$ are the discrete values corresponding to them. If for a record RID in R, $X \subseteq RID$, then we say that the record RID includes the quantitative itemset X.

The support of a quantitative itemset X in a relational database is a percentage of records including X. It is also represented as supp(X) like in a relational database.

Below definitions formulate a quantitative association rule and a rare quantitative association rule in a relational database.

**Definition 3**. Let $I=\{f_j = q_{jk}/\ j=1,\ ...,\ n;\ k=1,\ ...,\ j_m\}$ be a quantitative itemset in a relational database, $X, Y \subseteq I$ be quantitative itemsets, $X \cap Y = \emptyset$, supp(X)$\neq$0, and supp(Y)$\neq$0. Also, the thresholds: min_supp, min_conf and min_interest > 0 are given by users or experts. Then, the rule X→Y can be extracted as *a quantitative association rule* if

(1) supp(X∪Y) $\geq$ min_supp,

(2) |supp(X∪Y)−supp(X)supp(Y)| $\geq$ min_interest,

(3) conf(X→Y) = supp ( X∪Y) / supp (X) $\geq$ min_conf.

**Definition 4**. Let $I=\{f_j = q_{jk}/\ j=1,\ ...,\ n;\ k=1,\ ...,\ j_m\}$ be a quantitative itemset in a relational database, $X, Y \subseteq I$ be quantitative itemsets, $X \cap Y = \emptyset$, supp(X)$\neq$0, and supp(Y)$\neq$0. Also, the thresholds: min_supp, min_conf and min_interest > 0 are given by users or experts. Then, the rule X→Y can be extracted as *a rare quantitative association rule* if

(1) supp(X) $\leq$ min_supp, supp(Y) $\leq$ min_supp,

(2) |supp(X∪Y) −supp(X)supp(Y)| ≥ min_interest,

(3) Interest(X, Y) > 1,

(4) conf(X→Y) = supp ( X∪Y) / supp (X) ≥ min_conf.

Based on these definitions, the process of rare quantitative association rule mining consists of two steps:

**Step1.** Identify all infrequent itemsets of potential interest. That is, from condition(1) of definition 4, Z is an infrequent itemset if $\exists X, Y: X \cap Y = \emptyset$, $X \cup Y = Z$, $\forall x_k \in X, y_k \in Y$, $supp(x_k) \le min\_supp$, $supp(y_k) \le min\_supp$, and from condition(2) of definition 4, Z is an infrequent itemset of potential interest if $|supp(X \cup Y) - supp(X)supp(Y)| \ge min\_interest$.

**Step2.** Extract rules of interest from these itemsets. That is, they are restricted to positive rules by Interest(X, Y) > 1 (condition(3) of definition 4), and strong rules of interest are extracted by conf(X→Y) ≥ min_conf (condition(4) of definition 4).

## 4. Hashing algorithm for network intrusion detection

As above discussion, the network packets database is a relational database and intrusion data appear very rarely than regularities in it. Therefore, data mining based IDS, where rare quantitative association rules are extracted for intrusion patterns, can be constructed.

And, network audit databases face tens of attributes, of which the record length is not so long, and same. Besides, the number of infrequent items is little, so that it is possible to employ the method by hashing for finding supports of *k*-infrequent itemsets.

The following algorithm, when the database is scanned for the first time, identifies all 1-infrequent itemsets, and when the database is scanned for the second time, counts supports of all *k*-infrequent itemsets for each record in database by hashing. And then, it chooses positive infrequent itemsets of potential interest satisfying |supp(X∪Y)− supp(X)supp(Y)| ≥ min_interest and Interest(X, Y)>1, and finally, extracts rare associative rules of strong interest satisfying conf(X→Y) ≥ min_conf.

**[Algorithm]**

Input      R: Relational database;

           min_supp: minimum support; min_interest: minimum interest;

           min_conf: minimum confidence

Output     RAR: Rare association rules for R

(1)   Obtain the family of all 1- infrequent itemsets $NL_1$, scanning database R.

(2)   Address $= \varnothing$

    for $\forall r \in R$ {

        $rL_1$= family of 1- infrequent itemsets (r);    // $rL_1 \subseteq NL_1$

        for(k=2; $rL_{k-1} \neq \varnothing$; k++){

            $rC_k$=apriori_gen($rL_{k-1}$);

            for $\forall c \in rC_k$ {

                Address(k, c) = Hash(c);

                if   Address(k, c) $\notin$ Address {

                    Address = Address $\cup$ Address(k, c);

                    Value(k, c) = 1; }

                else

                   Value(k, c)++;

           } //Hashing

    } //support

   (3) NL $= \varnothing$

    for $\forall$Address(k, c) $\in$ Address {

        $NL_k$= {$Hash^{-1}$(Address(k, c))| Value(k, c) $\leq$ min_supp};

        NL=NL$\cup NL_k$;

    } //NL: All infrequent itemsets

   (4) RAR $= \varnothing$

$\forall X, Y \subseteq NL, X \cap Y = \emptyset,$

if $(|supp(X \cup Y) - supp(X)supp(Y)| \geq min\_interest) \wedge (Interest(X, Y) > 1) \wedge (supp(X \cup Y) / supp(X) \geq min\_conf)$

RAR = RAR $\cup$ { X$\rightarrow$Y }; //All Rare Association Rules

return RAR ;

Procedure apriori_gen($NL_{k-1}$)

$C_k = \emptyset$

for $\forall l_1 \in NL_{k-1}$ {

 for $\forall l_2 \in NL_{k-1}$ {

  if $(l_1[1]=l_2[1]) \wedge (l_1[2]=l_2[2]) \wedge \ldots \wedge (l_1[k-2]=l_2[k-2]) \wedge (l_1[k-1]<l_2[k-1])$ then {

  c=$l_1 \infty l_2$;

  $C_k = C_k \cup_k c$ ; }}}

return $C_k$;

In the following, we will give a concrete example, to illustrate the above algorithm design. Table 2 shows a part of network packets database.

Table2. Network packets database

| ID | service | src_bytes | dst_bytes | duration | ... |
|----|---------|-----------|-----------|----------|-----|
| r1 | telnet | 100 | 2000 | 13 | ... |
| r2 | ftp | 200 | 300 | 2 | ... |
| r3 | smtp | 250 | 300 | 1 | ... |
| r4 | telnet | 200 | 12100 | 60 | ... |
| r5 | smtp | 200 | 300 | 1 | ... |
| ... | ... | ... | ... | ... | ... |

In Table 2, {telnet, ftp, smtp} is a domain of discrete values for the field $f_1$:service, {dst_bytes ≤ 1000, dst_bytes > 1000} for $f_3$:dst_bytes, {duration ≤ 10, duration > 10} for $f_4$:duration, and so on. Then, by using these domains of the values, data in Table 2 can be transformed into Table 3.

Table 3. Discretization result of network packets database

| ID | service | src_bytes | dst_bytes | duration | … |
|----|---------|-----------|-----------|----------|----|
| r1 | A | D | E | G | … |
| r2 | B | D | F | H | … |
| r3 | C | D | F | H | … |
| r4 | A | D | E | G | … |
| r5 | C | D | F | H | … |
| … | … | … | … | … | … |

A set of all items in Table 3 is I={A, B, C, D, E, F, G, H}, where A: [$f_1$=telnet], B: [$f_1$=ftp], C: [$f_1$=smtp], D: [$f_2$= src_bytes ≤ 300], E: [$f_3$= dst_bytes > 1000], F: [$f_3$= dst_bytes ≤ 1000], G: [$f_4$= duration > 10], H: [$f_4$= duration ≤ 10].

Therefore, there are five records and eight items in this example. We suppose that min_supp = 0.5, min_interest = 0.05, and min_conf = 0.5 are given by users or experts. And then, we can apply algorithm as follows:

1) Obtain the family of all 1- infrequent itemsets, say NL1, scanning database R, by using min_supp = 0.5 (Figure 1).

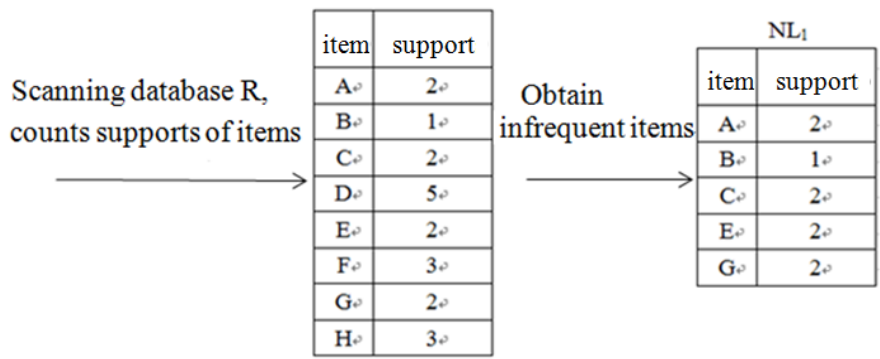

Figure 1. Identification of infrequent items in database R

2) For each record, identify infrequent items scanning database for the second time, generate as many combination of them as possible, i.e. $k$-infrequent itemsets (NL$_k$), and count their supports by hashing (Figure 2).

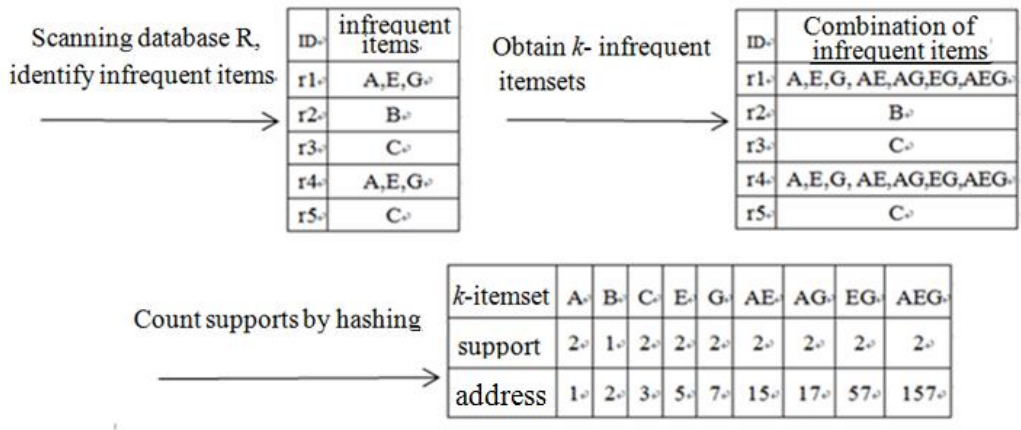

Figure 2. Counting *k*-infrequent itemsets and supports by hashing

3) Extract rare association rules of interest satisfying |supp(X∪Y)−supp(X)supp(Y)| > min_interest.

In this example, for X=A and Y=E, |supp(A∪E)−supp(A)supp(E)| = |0.4 − 0.4*0.4| = 0.24. This value is larger than min_interest=0.05, and A→E is extracted as a rare association rules of interest.

4) Choose positive association rules satisfying Interest(X, Y) = supp(X∪Y) / (supp(X)supp(Y)) > 1.

In this example, for A→E, Interest(A, E) = supp(A∪E) / (supp(A)supp(E)) = 0.4 / 0.4*0.4 = 2.5 > 1. Thus, A→E is extracted as a positive association rule.

5) Choose association rules of strong interest satisfying conf(X→Y) = supp ( X∪Y) / supp (X) ≥ min_conf.

In this example, for A→E, conf (A→E) = supp(A∪E) / supp(A) = 0.4 / 0.4 = 1.   This value is large than min_conf=0.5, and A→E is extracted as a association rules of strong interest.

Thus, we aimed at anomaly cases occurring rarely in IDS, which uses network audit database facing tens of attributes, and developed a rare association rule mining algorithm by hashing, which is very in practice, for finding intrusion patterns.

## 5. Conclusion

In this paper, we propose a new practical association rule mining algorithm for anomaly detection in Intrusion Detection System (IDS). First, with a view of anomaly cases being relatively rarely occurred in network packet database, we define a rare association rule among infrequent itemsets rather than the traditional association rule mining method. And then, we discuss an interest measure to catch differences between interesting relations and uninteresting ones, and what interest there is, and develop a hash based rare association rule mining algorithm for finding rare, but useful anomaly patterns to user. Finally, we define a quantitative association rule in relational database, propose a practical algorithm to mine rare association rules from network packet database, and show advantages of it giving a concrete example.

Our algorithm can be applied to fields need to mine hidden patterns which are rare, but valuable, like IDS, and it is based on hashing method among infrequent itemsets, so that it has obvious advantages of speed and memory space limitation problems over the traditional association rule mining algorithms.